\documentclass[12pt,preprint]{aastex}

\def\kpc{\,{\rm kpc}}
\def\phibar{\varphi_{\rm bar}}
\def\etal{{et al.}}
\def\eg{{\it e.g.}}

\def\ie{{\it i.e.}}

\def\cobe{{\it COBE}}
\def\cobedirbe{{\it COBE}/DIRBE}
\def\cober{{\it COBE-$\rho$}}
\def\cobed{{\it COBE}-Dyn}
\def\kms{$\mathrm {km}~\mathrm{s}^{-1}$}
\def\kmsk{$\mathrm {km}~\mathrm{s}^{-1}~ \mathrm{kpc}^{-1}$}
\def\pc{{\rm pc}}
\def\degrees{^\circ}
\def\mlo{M_L^<}
\def\mhi{M_L^>}
\def\msun{M_\odot}

\shorttitle{Large-Scale Model of the Milky Way}
\shortauthors{Bissantz \etal}

\begin{document}

\title{Large-Scale Model of the Milky Way: Stellar Kinematics and 
Microlensing Event Timescale Distribution in the Galactic Bulge}

\author{Nicolai~Bissantz\altaffilmark{1}, Victor P.~Debattista\altaffilmark{2} and Ortwin~Gerhard}
\affil{Astronomisches Institut, Universit\"at Basel, Venusstrasse 7,
CH-4102 Binningen, Switzerland}
\email{bissantz@math.uni-goettingen.de, debattis@phys.ethz.ch,
       gerhard@astro.unibas.ch}

\altaffiltext{1}{current address: Inst.\ for Mathem.~Stoch.,
Maschm\"uhlenweg 8-10, 37083 G\"ottingen, Germany}
\altaffiltext{2}{current address: Inst.\ f\"ur Astronomie, ETH
  H\"onggerberg, HPF G4.2, CH-8093, Z\"urich, Switzerland}

\begin{abstract}
  We build a stellar-dynamical model of the Milky Way barred bulge and
disk, using a newly implemented adaptive particle method. The
underlying mass model has been previously shown to match the Galactic
near-infrared surface brightness as well as gas-kinematic
observations. Here we show that the new stellar-dynamical model also
matches the observed stellar kinematics in several bulge fields, and
that its distribution of microlensing event timescales reproduces the
observed timescale distribution of the {\it MACHO} experiment with a
reasonable stellar mass function.  The model is therefore an excellent
basis for further studies of the Milky Way. We also predict the
observational consequences of this mass function for parallax shifted
events.
\end{abstract}

\section{Introduction}
It is now known, from several independent lines of evidence, that the
Milky Way Galaxy (MWG) is barred (e.g., \citet{ger01}).  However, a
comprehensive model consistent with the main observables -- luminosity
distribution, stellar-kinematics, gas-kinematics, and microlensing --
has so far been still missing.  Recently, \citet{bg02} obtained a
luminosity density model for the MWG from the dust-corrected $L$-band
\cobedirbe\ map of \citet{smb95}, through a non-parametric constrained
maximum likelihood deprojection.
This model (hereafter: \cober\ model) is consistent also with the
observed magnitude distributions of clump giant stars towards several
bulge fields, and with the microlensing optical depth towards the
bulge derived from these stars \citep{popo03, alf03}; see also
\citet{bbg00} and \citet{bg02}.  Furthermore
\cite{beg03} found that the hydrodynamical gas flow in the potential
of the \cober\ model matches the observed gas dynamics of the inner
MWG well.

The structure of the inner MWG can also be constrained by observations
of stellar kinematics along fixed lines of sight (\citet{swc90}
[Sh90]; \citet{sjw92} [Sp92]; \citet{metal92} [Mi92]) and by the
microlensing event timescale distribution (ETD) \citep{al00}.  The ETD
has been studied largely with models which assume some distribution of
disk and bulge kinematics (\eg\ \citet{hg96, peale98, mcs98}).  An
exception was \citet{zrs96}, who used the dynamical bar model of
\citet{zhao96} augmented by an analytic disk model, but failed to
match the long duration ($\hat t > 100$ days) tail of the ETD.
In the present Letter we show that a full stellar-dynamical model
based on the \cober\ model is consistent with these independent data
as well.

Dynamical models of the MWG have been generated using the
\citet{sch79} method, in which the distribution function (DF) of a
galaxy is built from numerically integrated stellar orbits.  Following
earlier work by \citet{zhao96}, \citet{hedb00} constructed a
22168-orbit dynamical model of the MWG.  Dynamical models of the MWG
have also been obtained by $N$-body methods \citep{fux97}.
\citet{st96} [ST96] introduced a novel method for generating
self-consistent dynamical models.  The Syer-Tremaine (hereafter ST)
method is allied to the Schwarzschild method, but, rather than
superposing time-averaged observables from an orbit library, the ST
method constructs a model by actively varying the weights of
individual particles (orbits) as a function of time.  This permits
arbitrary geometry and a larger number of orbits to be used in the
model building. Our dynamical model for the \cober\ density in the MWG
is constructed with the ST method, demonstrating its usefulness for
real galaxy modeling.  This Letter compares the model's bulge
kinematics and microlensing ETD with their observed counterparts.

\section{The Syer-Tremaine method}
\label{SYTR}

The idea of the ST algorithm is to assign individual weights to particles of a
simulation, which are then changed to reduce the deviation between the model
and observations.  An observable $Y_j$ associated to a stellar system 
characterized by a distribution function $f({\bf z})$, $\bf
z=(x,v)$, can be written as $Y_j=\int K_j({\bf z}) f({\bf z}) d^6{\bf z}$,
where $K_j({\bf z})$ is a known kernel.  If this stellar system is simulated
with $N$ particles having weights
$w_i$ and phase space coordinates $\bf z_i$, then we can write the observables
of the simulation as $y_j(t)=\sum_{i=1}^N w_i(t) K_j({\bf z_i}(t))$.  ST96
define the "force of change" on the weights as
\begin{equation}
\frac{dw_i(t)}{dt}=-\varepsilon w_i(t) \sum_{j}
\frac{K_j({\bf z_i})}{Z_j} \left(\frac{y_j(t)}{Y_j}-1 \right)
\end{equation}
The small and positive parameter $\varepsilon$ governs how rapidly the
weights are pushed such that the simulation observables $y_j(t)$
converge towards the observables $Y_j$.  The constants $Z_j$ act as
normalizations.  The full ST method also includes prescriptions for
temporal smoothing and a maximum entropy term, to reduce fluctuations.
We have implemented the ST method with the MWG disk-plane surface
density as the observable (Debattista \etal\ in preparation).  We set
$\varepsilon = 0.25$, $\alpha = 0.524$, $\mu = 0.001$, where $\alpha$
and $\mu$ are the parameters of the temporal smoothing and the entropy
terms, respectively, in the notation of ST96.

\subsection{Simulation}

Since the MWG contains a bar, our initial model also had to be barred.
The simplest way to achieve this was to evolve an $N$-body model of an
initially axisymmetric, bar-unstable, disk galaxy.  The $N$-body
simulation which produced the barred model consisted of live disk and
bulge components inside a frozen halo.  The frozen halo was
represented by a cored logarithmic potential.  The initially
axisymmetric disk was modeled by a truncated exponential disk.  Disk
kinematics were set up using the epicyclic approximation to give
Toomre $Q = 1.3$.  The disk and bulge were represented by $4\times
10^6$ equal-mass particles, with a mass ratio $M_{\rm d}:M_{\rm b} =
4:1$.  Further details of the setup methods and model units can be
found in \citet{d03} [D03].  We use the halo, disk, and bulge parameters
given in Table 2 of D03, which give a flat rotation curve
out to large radii.

The simulation was run on a 3-D cylindrical polar grid code (described
in  \citet{sv97}), with technical parameters
exactly as in D03.  The initially axisymmetric system was
unstable and formed a rapidly rotating bar at $t \simeq 50$.  By
$t=160$, the bar instability had run its course and further secular
evolution of the bar was mild.  The resulting system did not match the
\cober\ model of the MWG and needed to be evolved further with the ST
code.  First, however, we eightfold symmetrized the \cober\ model in
order to reduce the amplitude of spirals, which we did not try to
reproduce.  We evolved the $N$-body model from $t=160$ under the ST
prescription with the fixed potential of the \cober\ model plus dark
matter halo. We kept the bar pattern speed
at its value in the $N$-body model, which scales to $56$ \kmsk,
consistent with the MWG (\citet{dehnen00}; \citet{dgs02};
\citet{beg03}).  At $t=240$ (\ie\ $\simeq 4$ bar rotations), we shut
off the ST algorithm and evolved the system to $t=280$ to assure that
the particles are phase mixed.

\section{Results: Density and Bulge Kinematics}

To compare our dynamical \cobe\ model (the \cobed\ model) with observations,
we adopted the same viewing parameters as were used to determine the \cober\ 
model: $R_\odot=8\kpc, z_{\odot}=14\pc$ and $\phibar=20\degrees$ \citep{bg02}.  
We scale the velocities in the \cobed\ model to the MWG by
matching to the local circular velocity.  We assumed that the local
standard of rest has only a circular motion, with $v_{LSR} = 200$ \kms, and we
adopted the values of the solar peculiar motion from \cite{db98}.

The densities of the \cobed\ and \cober\ models match very well, with
azimuthally averaged errors smaller than $5\%$ out to $R_\odot$.  The largest
errors ($< 15\%$) occur in small isolated regions on the bar major-axis.  In
the (unconstrained) vertical direction, the disk is somewhat thicker than the
MWG at $R_\odot$, but this leads to a change in optical depth 
$\tau$ towards Baade's
window of $<15\%$.  In the bulge region, on the other hand, the scale-height
of the \cobed\ model matches that of the MWG very well.
We compared the model's kinematics to observations towards Baade's window
(Sh90, Sp92) and in the field at $(l,b)=(8\degrees,7\degrees)$ (Mi92), using
the selection functions determined by \citet{hedb00}.  Table
\ref{tab1} shows our results.  The overall fit of our model to the observed
kinematics is rather good.

\section{The Microlensing ETD}
\label{micromethod}

We now show that the \cobed\ model is also consistent with the
microlensing ETD.  \citet{al00} presented an ETD, corrected for their
experimental detection efficiency, based on 99 events in 8 fields.
\citet{popo02} argued that one of these fields seems biased towards
long-duration events, introducing some uncertainty in the observed ETD.  Here
we use the full-sample Alcock \etal\ ETD in order that our results may be
compared with previous ones.
We computed the ETD with the self-consistent kinematics of the \cobed\ model.
A microlensing event is characterized by the source distance, $D_S$, lens
distance, $D_L$, the proper motion, $v_\perp$, of the lens with respect to the
line-of-sight between observer and source, and lens mass, $M_L$.  The
probability $P(\hat{t})$ for observing an event duration 
$\hat t=2 \Theta_E/v_\perp$ is given by
\begin{eqnarray}
P(\hat{t}) \propto
\int
\rho(D_S) D_S^{2+2\beta} \rho(D_L) D_L^2  \Theta_E(D_S,D_L,M_L) &&\nonumber \\
\Phi(M_L) v_\perp f(v_\perp) \delta(\hat t- 2 \Theta_E/v_\perp) dv_\perp~ dD_L~
dD_S~ dM_L. &&
\label{eqn:etd}
\end{eqnarray}
Here $\rho(d)$ is the density of the MWG at distance $d$ from the
observer along the line-of-sight to the observed field, $\Phi(M_L)$ is
the mass function (MF) of the lens population, $\Theta_E(D_S,D_L,M_L)$
is the Einstein angle, and $f(v_\perp)$ is the distribution of
$v_\perp$.  We solved the multiple integral by Monte Carlo random
drawings of the parameters $(D_S, D_L, v_\perp, M_L)$ as follows: (1)
To obtain the {\bf source distance} ($0 \leq D_S \leq D_S^{\rm max}=
12\kpc$), we used the \cober\ model, since this is less noisy than the
particle realization.  The probability of $D_S$ is $\propto
\rho(D_S)D_S^{2+2\beta}$ with $\beta\!=\!-1$, 
to account for a magnitude cut-off \citep{kp94}.  (2)
The {\bf lens distance} ($0\leq D_L < D_S$) was selected from
$\tilde{\rho}(D_L) \int_0^{D_S} \rho(D_L) dD_L$, where $\tilde{\rho}$
is a normalized probability density distributed as in the \cober\ 
model.  (3) For the {\bf relative velocity $v_\perp$}, we used the
particle distribution of the \cobed\ model, randomly selecting a
particle at $\sim D_S$ and another at $\sim D_L$.  The proper motions
of these particles then determined $v_\perp$.  (4) The {\bf lens mass
  $M_L/M_{\odot}$} was selected from a \citet{kroupa95} MF, $\Phi(M_L)
= \beta (M_L/\msun)^{-\gamma}$, with
\begin{equation}
(\gamma,\beta) = 
\cases{ \displaystyle 
 (2.35,0.1038) & $ \mlo \leq M_L/M_{\odot} \leq 0.35$ \cr
  (0.6,0.6529) & $ 0.35 \leq M_L/M_{\odot} \leq  0.6$ \cr
 (2.35,0.2674) & $  0.6 \leq M_L/M_{\odot} \leq \mhi$ \cr}
\end{equation}
We explored varying $\mlo$ and $\mhi$.  We obtained the ETD, shown in
Fig. \ref{fig1}, by simulating $10^5$ events, and weighting each by
the remaining factors in Eqn.  \ref{eqn:etd}.  We tested our Monte
Carlo integrations by reproducing one of \citet{peale98}'s model ETDs.

We started with $(\mlo,\mhi) = (0.075,10)$, for which we obtained a
Kolmogorov-Smirnov distance between data and model of $D_{KS} =
0.213$.  (We excluded the bin at $\hat{t} < 3.1$ days from the {\it
  MACHO} data in all such comparisons, because it appears to be
too heavily affected by its large detection-efficiency correction.)
To improve on this fit, we first explored the effects of uncertainties
in the \cober\ model.  The most important of these is $\phibar$.
Setting $\phibar = 30\degrees$, we found only a minor change to the
ETD, in agreement with \citet{peale98}.  
Making the bar stronger, or the disk velocity dispersion outside the bar
smaller did not alter the ETD substantially.  Therefore we next
explored variations in the MF.  Like \citet{peale98}, we found that
modest changes can improve the fit substantially.  Our best fit, with
$D_{KS} = 0.068$ was obtained with $\mlo = 0.04$ and $\mhi = 10$.
However, a more conservative limit is $\mhi = 4$, which gives $D_{KS}
= 0.081$.  (If the suggestion of \citet{popo02} is correct, which
would shift the ETD peak to smaller $\hat{t}$, then a smaller $\mhi$
would be required anyway.)

We now explore the causes of long duration (LD) events in the \cobed\
model; using $(\mlo,\mhi) = (0.04,4)$ as our standard model for this
analysis.  We start by noting, from Fig.  \ref{fig1}, that the vast
majority of sources are located in the bulge ($6 \leq D_S \leq 10$
kpc).  This is also true for the lenses responsible for short duration
events, but disk lenses become more important at longer durations;
indeed, for ${\hat t} > 25$ days, one third of the lenses are at $D_L
< 4$ kpc.  In Fig. \ref{fig2} we separate the ETD into the near and
distant lens sub-samples and show the heliocentric angular velocities
and cumulative distributions of $M_L$ for both.  Note first that
lenses with $M_L > 0.5 \msun$ contribute significantly to LD events in
both the near and distant sub-samples.  Lens mass, however, is not the
full explanation of the LD events, as has been noted by previous
studies, and the relative motions of lens and source in the
heliocentric frame must also be considered.  The kinematics of the LD
sources are substantially those of a rotating triaxial bulge/bar which
points almost towards the observer: thus their apparent tangential
motions are largely due to the solar motion, giving $\Omega_{tan,S}
\sim 205/8$ \kmsk.  Distant lens LD events are then possible because
the lenses share very similar kinematics with the sources (note that
massive lenses become necessary only in the last quartile, $\hat{t} >
60$ days).  For the nearby lens sample, LD events have a rather large
spread in $\Omega_{tan,L}$ (due both to their proximity and the
velocity dispersion of the \cobed\ model), which together with larger
$M_L$'s is able to produce LD events. We conclude, therefore, that there
is no single cause for the long-duration events.

Standard 3-parameter fits to microlensing events are symmetric about
the time of peak amplification, resulting in a degeneracy between
$M_L$, $v_\perp$, $D_L$ and $D_S$.  One degree of degeneracy is
removed by also measuring the light-curve shift due to the parallax
from earth's orbit, which gives a relation between $v_\perp$ and
$D_L/D_S$.  These shifts are present in all events, but most go
undetected because of infrequent sampling and photometric errors.
\citet{bk97} estimate a $1\%$ detection efficiency of parallax-shifted
events for the {\it MACHO}-type setup, and much higher for
second generation experiments.  The light curves of such events require 5
parameters, including $\kappa \equiv R_\oplus(D_L^{-1} -
D_S^{-1})/\Theta_E$, where $R_\oplus = 1$ AU.  In Fig.  \ref{fig3}, we
present our predictions for the probability distribution in the
$(\kappa,\hat{t})$-plane, assuming $100\%$ detection efficiency.
These distributions are twin-peaked, with the lower peak increasingly
separated from the global peak as $\mlo$ decreases, as it must since
$\kappa \propto \Theta_E^{-1}$ while $\hat{t} \propto \Theta_E$.  The
location of the second peak may therefore provide an observational
constraint on the MF at low mass.

\section{Conclusions}
\label{sectconcl}

We have presented a dynamical model of the MWG constructed using the
Syer-Tremaine method, constrained only by the MWG density map of
\citet{bg02}.  Although no kinematic constraints were used, the model
(i) matches observed bulge kinematics in several fields and is (ii)
able to reproduce the observed microlensing event timescale
distribution.  For the best fitting MF, the model (iii) predicts a
twin-peaked probability distribution in the $(\kappa,\hat{t})$-plane,
which may be observationally tested with new generations of
microlensing experiments.  (iv) The underlying mass model has been
previously shown to match the Galactic near-infrared surface
brightness as well as gas-kinematic observations. It is therefore an
excellent basis for further studies of the Milky Way.

\acknowledgments
\noindent
This work was supported by the Schweizerischer Nationalfonds through grant
20-64856.01.

\clearpage

\begin{table}[!ht]
\begin{centering}
\begin{small}
\begin{tabular}{c|cccc}
\hline
 & $(l,b)$ & Ref. & Observed & \cobed  \\
\hline
$v^h_{\rm los}   $ & $(1\degrees,-4\degrees)$ & Sh90 & $4\pm8$     & 8.2 \\
$\sigma_{\rm los}$ & $(1\degrees,-4\degrees)$ & Sh90 & $113\pm5$   & 109 \\
$\sigma_{\rm los}$ & $(1\degrees,-4\degrees)$ & Sp92 & $120$       & 109 \\
$\sigma^h_{\mu_l}$ & $(1\degrees,-4\degrees)$ & Sp92 & $3.2\pm0.1$ & 3.1 \\
$\sigma^h_{\mu_b}$ & $(1\degrees,-4\degrees)$ & Sp92 & $2.8\pm0.1$ & 2.4 \\
$v^g_{\rm los}   $ & $(8\degrees, 7\degrees)$ & Mi92 & $45\pm10$   & 46  \\
$\sigma_{\rm los}$ & $(8\degrees, 7\degrees)$ & Mi92 & $85\pm7$    & 100 \\ 
\hline
\end{tabular}
\caption{Comparison of kinematic quantities computed from the \cobed\
  model with observations. In the first column, the superscript $^h$
  indicates that the value given is heliocentric, and $^g$ that it is
  Galactocentric.  All quantities are in \kms\ except for $\sigma^h_{\mu_l}$
  and $\sigma^h_{\mu_b}$, which are in milliarcseconds year$^{-1}.$}
\label{tab1}
\end{small}
\end{centering}
\end{table}

\clearpage

\noindent {\bf Figure Captions}

\noindent{\bf Fig.~1}
The ETD of the \cobed\ model compared to the
detection-efficiency-corrected observations of the {\it MACHO} group
(histograms in all panels).  Top: cumulative distribution function for
the standard model, $(\mlo,\mhi) = (0.04,4)$, (solid line), the best
model with $(\mlo,\mhi) = (0.04,10)$ (dotted), and a model with
$(\mlo,\mhi) = (0.075,10)$ (dashed).  We obtain $D_{KS} = 0.081,
0.068, 0.213$, respectively, for the 3 models.  Middle panel:
differential distributions of these models (same line styles).  (In
these two panels, all model distributions have been smoothed with a
kernel density estimator of bandwidth $0.1$.)  Bottom: ETD of the
$(\mlo,\mhi) = (0.04,4)$ model and its decomposition into events with:
$6 \leq D_S \leq 10$ kpc (dotted line), $0 \leq D_L \leq 4$ kpc
(dashed) and $D_L > 4$ kpc (dot-dashed).

\noindent{\bf Fig.~2}
Top two panels: The long duration events ($\hat{t} > 42$ days) in the
$(\mlo,\mhi) = (0.04,4)$ model.  Both are maps (on the same relative
scale) in the plane of heliocentric tangential angular velocities,
$\Omega_{tan,L}$ and $\Omega_{tan,S}$.  Left: near lenses ($D_L < 4$
kpc); right: distant ones ($D_L > 4$ kpc).  The diagonal and
horizontal dashed lines indicate $\Omega_{tan,L} = \Omega_{tan,S}$ and
$\Omega_{tan,S} = 205/8 = 25.6$ \kmsk, respectively.  Bottom:
distributions of $M_L$ for distant (solid lines) and nearby (dashed)
lenses.  The different lines result from splitting into quartiles by
contribution to the full ETD the distribution of events sorted on
$\hat{t}$.  Event durations increase as the mean mass increases.  The
heavy curve shows the underlying mass function.

\noindent{\bf Fig.~3} 
  Predicted probability distribution of parallax shifted events
  in the $(\kappa,\hat{t})$-plane, for the standard model with
  $(\mlo,\mhi) = (0.04,4)$.  We use a smoothing kernel with
  $(\delta_\kappa,\delta_{\log\hat{t}}) = (0.01,0.1)$.  The stars mark
  the locations of secondary peaks when $\mlo = $ 0.075, 0.04, 0.02 and
  0.01 in order of increasing $\kappa$.

\clearpage

\begin{figure}
\plotone{f1.eps}
\caption{ }
\label{fig1}
\end{figure}

\begin{figure}
\plotone{f2.eps}
\caption{ }
\label{fig2}
\end{figure}

\begin{figure}
  \plotone{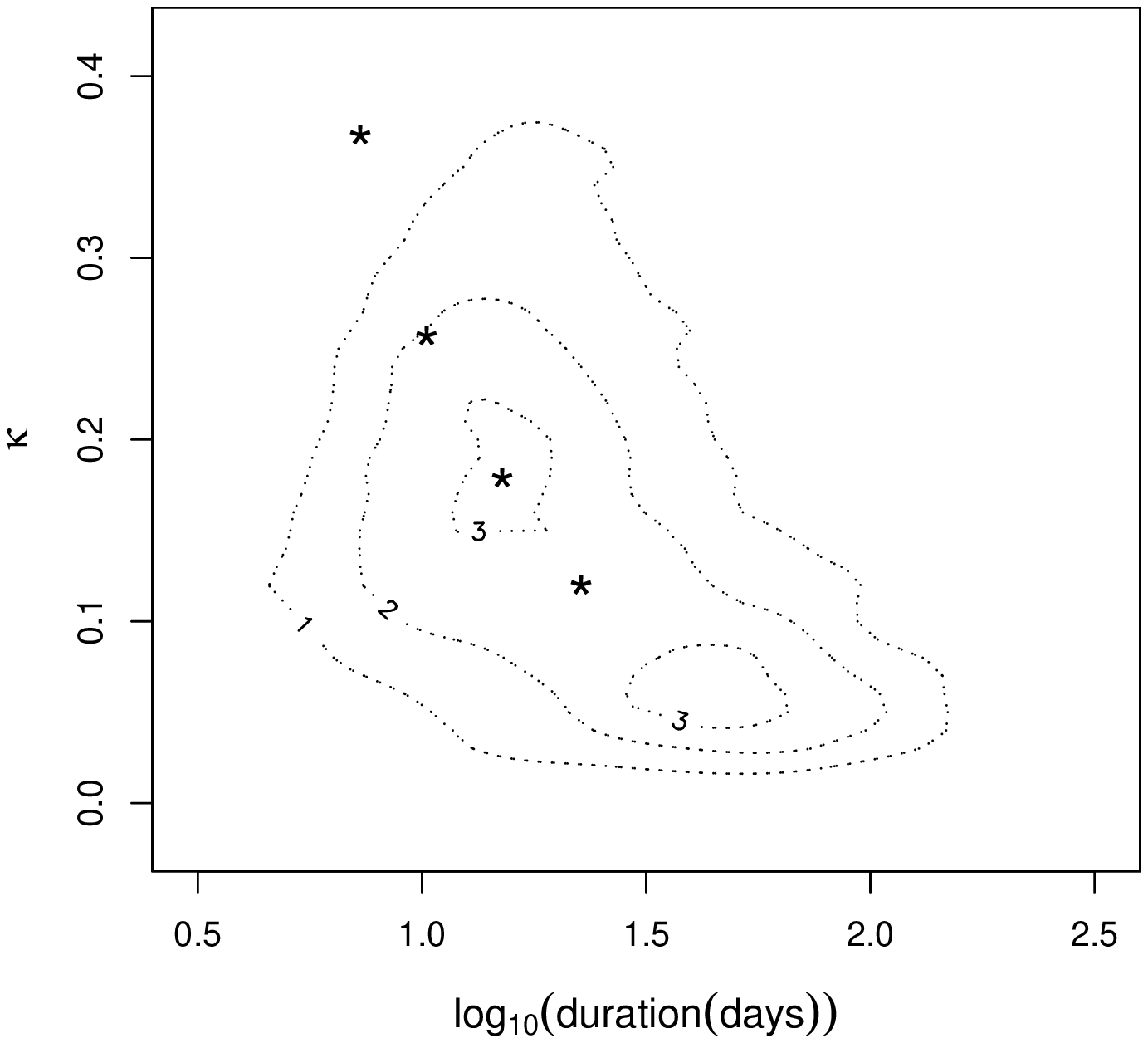}
\caption{ }
\label{fig3}
\end{figure}

\clearpage 

\end{document}